\documentclass[12pt]{article}
\usepackage{graphicx}
\def\sech{\mathop{\mathrm{sech}}}
\def\tanh{\mathop{\mathrm{tanh}}}
\def\Arctanh{\mathop{\mathrm{Arctanh}}}

\def\sinh{\mathop{\mathrm{sinh}}}
\def\cosh{\mathop{\mathrm{cosh}}}
\def\cn{\mathop{\mathrm{cn}}}

\def\C{{\rm\kern.24em \vrule width.02em 
 height1.4ex depth-.05ex
\kern-.26em C}}
\def\R{{\rm I\kern-.20em R}}
\def\Z{{\rm\kern.26em \vrule width.02em height0.5ex depth0ex
 \kern.04em
 \vrule  width.02em height1.47ex depth-1ex \kern-.34em Z}}
\def\N{{\rm I\kern-.20em N}}
\def\Q{{\rm\kern.24em \vrule width.02em 
 height1.4ex depth-.05ex \kern-.26em Q}}
\begin{document}
\begin{center}
{\bf \Large Shallow Water Waves and Solitary Waves}
\vskip 8pt
{\sc Willy Hereman} 
\vskip 5pt
Department of Mathematical and Computer Sciences, \\
Colorado School of Mines, \\
Golden, Colorado, USA 
\end{center}
%
%
%
\section*{Article Outline}
\noindent
Glossary
\vskip 4pt
\noindent
I.$\;\;$ Definition of the Subject
\vskip 3pt
\noindent
II.$\;\;$Introduction--Historical Perspective
\vskip 3pt
\noindent
III.$\;\;$Completely Integrable Shallow Water Wave Equations
\vskip 3pt
\noindent
IV.$\;\;$Shallow Water Wave Equations of Geophysical Fluid Dynamics
\vskip 3pt
\noindent
V.$\;\;$Computation of Solitary Wave Solutions
\vskip 3pt
\noindent
VI.$\;\;$Water Wave Experiments and Observations
\vskip 3pt
\noindent
VII.$\;\;$Future Directions 
\vskip 3pt
\noindent
VIII.$\;\;$Bibliography
%
%
\section*{Glossary}
\noindent
{\sc Deep water}
\vskip 2pt
\noindent
A surface wave is said to be in deep water if its wavelength is much shorter 
than the local water depth.
\vskip 7pt
\noindent
{\sc Internal wave}
\vskip 2pt
\noindent
A internal wave travels within the interior of a fluid.
The maximum velocity and maximum amplitude occur within the fluid or at an 
internal boundary (interface).
Internal waves depend on the density-stratification of the fluid.
\vskip 7pt
\noindent
{\sc Shallow water}
\vskip 2pt
\noindent
A surface wave is said to be in shallow water if its wavelength is much  
larger than the local water depth.
\vskip 7pt
\noindent
{\sc Shallow water waves}
\vskip 2pt
\noindent
Shallow water waves correspond to the flow at the free surface of a body of 
shallow water under the force of gravity, or to the flow below a horizontal 
pressure surface in a fluid.
%
%
\vskip 7pt
\noindent
{\sc Shallow water wave equations}
\vskip 2pt
\noindent
Shallow water wave equations are a set of partial differential equations that 
describe shallow water waves.
\vskip 7pt
\noindent
{\sc Solitary wave}
\vskip 2pt
\noindent
A solitary wave is a localized gravity wave that maintains its coherence and, 
hence, its visibility through properties of nonlinear hydrodynamics.  
Solitary waves have finite amplitude and propagate with constant speed and 
constant shape. 
\vskip 7pt
\noindent
{\sc Soliton}
\vskip 2pt
\noindent
Solitons are solitary waves that have an elastic scattering property: 
they retain their shape and speed after colliding with each other. 
\vskip 7pt
\noindent
{\sc Surface wave}
\vskip 2pt
\noindent
A surface wave travels at the free surface of a fluid.
The maximum velocity of the wave and the maximum displacement of fluid 
particles occur at the free surface of the fluid.
\vskip 7pt
\noindent
{\sc Tsunami}
\vskip 2pt
\noindent
A tsunami is a very long ocean wave caused by an underwater earthquake, 
a submarine volcanic eruption, or by a landslide.
%
\vskip 7pt
\noindent
{\sc Wave dispersion}
\vskip 2pt
\noindent
Wave dispersion in water waves refers to the property that longer 
waves have lower frequencies and travel faster. 
%
%
\section*{I.$\;$Definition of the Subject}
%
The most familiar water waves are waves at the beach caused by wind or tides, 
waves created by throwing a stone in a pond, by the wake of a ship, 
or by raindrops in a river (see Figure~{\ref{sww-fig1-raindrops1}). 
Despite their familiarity, these are all different types of water waves.
This article only addresses shallow water waves, 
where the depth of the water is much smaller than the wavelength of the 
disturbance of the free surface. 
Furthermore, the discussion is focused on gravity waves in which buoyancy 
acts as the restoring force. 
Little attention will we paid to capillary effects, and capillary waves for 
which the primary restoring force is surface tension are not covered.

\begin{figure}[htb]
\begin{center}
\includegraphics[width=3.73in, height=3.0in]{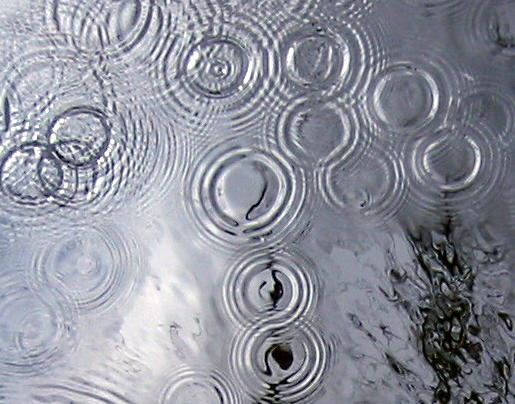}
\end{center}
\caption{Capillary surface waves from raindrops. 
Photograph courtesy of E.\ Scheller and K.\ Socha.
Source: MAA Monthly {\bf 114}, pp.\ 202-216, 2007.} 
\label{sww-fig1-raindrops1}
\end{figure}

Although the history of shallow water waves (Bullough 1988, Craik 2004, 
Darrigol 2003) goes back to French and British mathematicians of 
the eighteenth and early nineteenth centuries, Stokes (1847) is 
considered one of the pioneers of hydrodynamics (see Craik 2005).
%
%
He carefully derived the equations for the motion of incompressible, 
inviscid fluid, subject to a constant vertical gravitational force, 
where the fluid is bounded below by an impermeable bottom and above by a 
free surface.
Starting from these fundamental equations and by making further simplifying 
assumptions, various shallow water wave models can be derived.
These shallow water models are widely used in oceanography and atmospheric 
science.

This article discusses shallow water wave equations commonly used in 
oceanography and atmospheric science.
They fall into two major categories: 
Shallow water wave models with wave dispersion are discussed in Section III.
Most of these are completely integrable equations that admit smooth solitary 
and cnoidal wave solutions for which computational procedures are outlined in 
Section V.
Section IV covers classical shallow water wave models without dispersion.
Such hyperbolic systems can admit shocks.
Section VI addresses a few experiments and observations. 
The article concludes with future directions in Section VII.

%
%
%
\section*{II.$\;$Introduction--Historical Perspective}
\label{introduction}
%

The initial observation of a solitary wave in shallow water was made by 
John Scott Russell, shown in Figure~{\ref{sww-fig2-jsrussell}}. 
Russell was a Scottish engineer and naval architect who was conducting 
experiments for the Union Canal Company to design a more efficient canal boat. 

\begin{figure}[htb]
\begin{center}
\includegraphics[width=3.00in, height=3.0in]{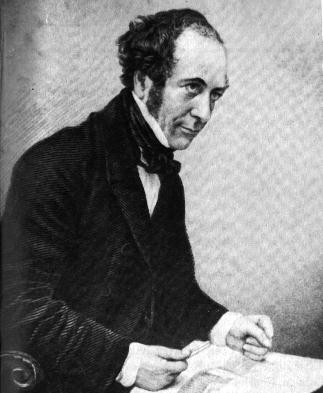}
\end{center}
\caption{John Scott Russell. Source: G.\ S.\ Emmerson (1977). 
Courtesy of John Murray Publishers.}
\label{sww-fig2-jsrussell}
\end{figure}

In Russell's (1844) own words:
``I was observing the motion of a boat which was rapidly drawn along a
narrow channel by a pair of horses, when the boat suddenly stopped--not so 
the mass of water in the channel which it had put in motion; 
it accumulated round the prow of the vessel in a state of violent agitation,
then suddenly leaving it behind, rolled forward with great velocity, 
assuming the form of a large solitary elevation, a rounded, smooth and 
well-defined heap of water, which continued its course along the channel
apparently without change of form or diminution of speed.
I followed it on horseback, and overtook it still rolling on at a rate
of some eight or nine miles an hour, preserving its original figure some
thirty feet long and a foot to a foot and a half in height.
Its height gradually diminished, and after a chase of one or two miles 
I lost it in the windings of the channel.
Such, in the month of August 1834, was my first chance interview with that
singular and beautiful phenomenon which I have called the Wave of 
Translation."
\vfill
\newpage
Russell built a water tank to replicate the phenomenon and research 
the properties of the solitary wave he had observed. 
Details can be found in a biography of John Scott Russell (1808-1882)
by Emmerson (1977), 
and in review articles by Bullough (1988), Craik (2004), and Darrigol (2003), 
who pay tribute to Russell's research of water waves.

In 1895, the Dutch professor Diederik Korteweg and his doctoral student 
Gustav de Vries (1895) derived a partial differential equation (PDE) 
which models the solitary wave that Russell had observed.
Parenthetically, the equation which now bears their names had already appeared 
in seminal work on water waves published by Boussinesq (1872, 1877) and 
Rayleigh (1876).
%
%
The solitary wave was considered a relatively unimportant curiosity in
the field of nonlinear waves.
That all changed in 1965, when Zabusky and Kruskal realized that the KdV 
equation arises as the continuum limit of a one dimensional anharmonic 
lattice used by Fermi, Pasta, and Ulam (1955) to investigate 
``thermalization" -- or how energy is distributed among the many possible 
oscillations in the lattice. 
Zabusky and Kruskal (1965) simulated the collision of solitary waves 
in a nonlinear crystal lattice and observed that they retain their shapes 
and speed after collision.
Interacting solitary waves merely experience a phase shift, advancing the 
faster and retarding the slower.
In analogy with colliding particles, they coined the word ``solitons" 
to describe these elastically colliding waves.
A narrative of the discovery of solitons can be found in Zabusky (2005).
%

Since the 1970s, the KdV equation and other equations that admit solitary 
wave and soliton solutions have been the subject of intense study 
(see, e.g., Remoissenet 1999, Filippov 2000, and Dauxois and Peyrard 2006).
Indeed, scientists remain intrigued by the physical properties and elegant 
mathematical theory of the shallow water wave models.
In particular, the so-called completely integrable models which can be solved 
with the Inverse Scattering Transform (IST). 
For details about the IST method the reader is referred to 
Ablowitz {\it et al} (1974), Ablowitz and Segur (1981), and
Ablowitz and Clarkson (1991). 
The completely integrable models discussed in the next section are 
infinite-dimensional bi-Hamiltonian systems, with infinitely many conservation 
laws and higher-order symmetries, and admit soliton solutions of any order.

As an aside, in 1995, scientists gathered at Heriot-Watt University for a 
conference and successfully recreated a solitary wave but of smaller 
dimensions than the one observed by Russell 161 years earlier 
(see Figure~{\ref{sww-fig3-recreation}}).
\vskip 0.60cm
\noindent
\begin{figure}[htb]
\begin{center}
\includegraphics[width=4.5in, height=3.0in]{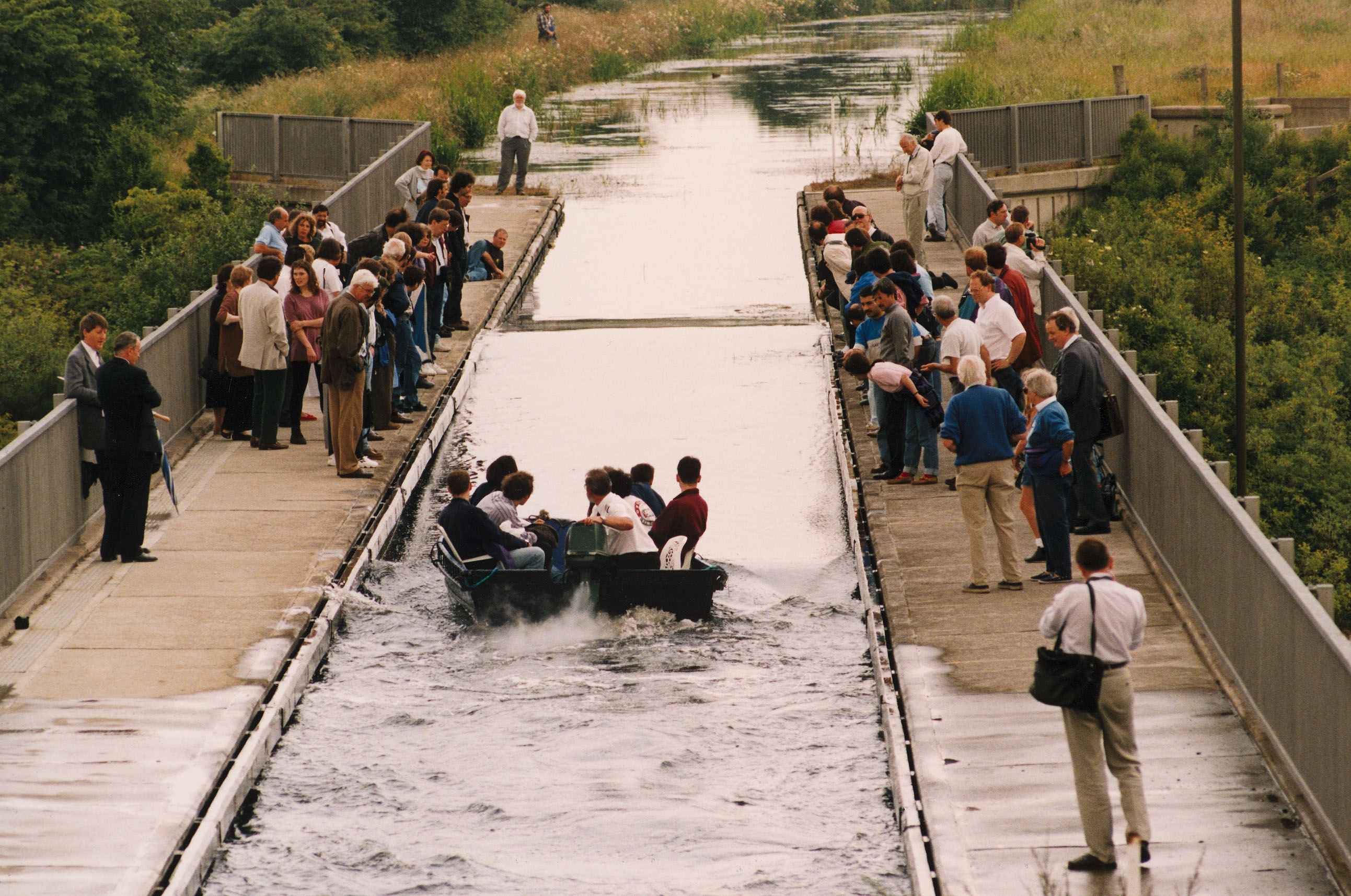}
\end{center}
\caption{Recreation of a solitary wave on the Scott Russell Aqueduct on 
the Union Canal. 
Photograph courtesy of Heriot-Watt University.}
\label{sww-fig3-recreation}
\end{figure}

%
%
%
%
%
%
\section*{III.$\;$Completely Integrable Shallow Water Wave Equations}
\label{dintegrableswwequations}
Starting from Stokes' (1847) governing equations for water waves, 
completely integrable PDEs arise at various levels of approximation in 
shallow water wave theory.
Four length scales play a crucial role in their derivation.
As shown in Figure~{\ref{sww-fig4-testwave2}}, the wavelength $\lambda$ of 
the wave measures the distance between two successive peaks. 
The amplitude $a$ measures the height of the wave, which is the
varying distance between the undisturbed water to the peak of the wave.
The depth of the water $h$ is measured from the (flat) bottom of the water 
up to the quiescent free surface.
The fourth length scale is along the $Y-$axis which is along the crest of 
the wave and perpendicular to the $(X,Z)-$plane. 

\begin{figure}[htb]

\begin{center}
%
%
\includegraphics[width=4.0in, height=3.0in]{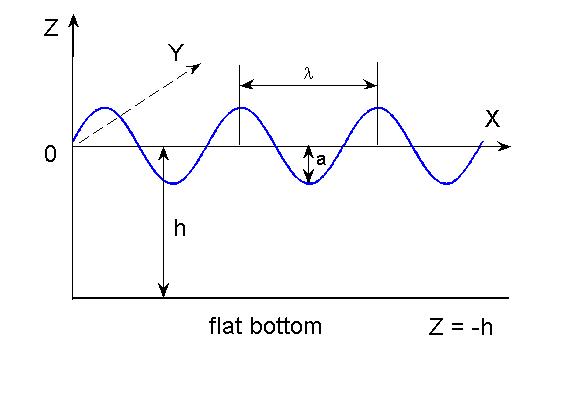}
\end{center}
\vspace{-15mm}
\caption{Coordinate frame and periodic wave on the surface of water.}
\label{sww-fig4-testwave2}
\end{figure}

Assuming wave propagation in water of uniform (shallow) depth, 
i.e., $h$ is constant, and ignoring dissipation, the model equations 
discussed in this section have a set of common features and limitations 
which make them mathematically tractable (Segur 2007b).
They describe 
(i) long waves (or shallow water), i.e., $h << \lambda,$
(ii) with relatively small amplitude, i.e., $a << h,$ 
(iii) travelling in one direction (along the $X-$axis) or weakly 
two-dimensional (with a small component in the $Y-$direction). 
Furthermore, the small effects must be comparable in size. 
For example, in the derivation of the KdV and Boussinesq equations 
one assumes that $\epsilon = a/h = O(h^2/\lambda^2),$
where $\epsilon$ is a small parameter $(\epsilon << 1),$ and 
$O$ indicates the order of magnitude.

\subsection*{The Korteweg-de Vries Equation}
The KdV equation was originally derived to describe shallow water waves
of long wavelength and small amplitude. 
In the derivation, Korteweg and de Vries assumed that all motion is uniform 
in the $Y-$direction, along the crest of the wave. 
In that case, the surface elevation (above the equilibrium level $h)$ of the 
wave, propagating in the $X-$direction, is a function only of the horizontal 
position $X$ (along the canal) and of time $T,$ i.e., $Z = \eta(X,T).$ 

%
In terms of the physical parameters, the KdV equation reads
\begin{equation}
\label{kdvoriginal}
\frac{\partial \eta}{\partial T} 
+ \sqrt{gh} \frac{\partial \eta}{\partial X}
+ \frac{3}{2}\frac{\sqrt{gh}}{h} \eta \frac{\partial \eta}{\partial X} 
+ \frac{1}{2} h^2 \sqrt{gh} (\frac{1}{3} - \frac{{\cal{T}}}{\rho g h^2}) 
\frac{\partial^3 \eta}{\partial X^3} = 0,  
\end{equation}
where $h$ is the uniform water depth, $g$ is the gravitational acceleration 
(about $9.81 {\rm m}/{\rm sec}^2$ at sea level), 
$\rho$ is the density, and ${\cal{T}}$ stands for the surface tension.
The dimensionless parameter 
${\cal{T}}/\rho g h^2$ 
is called the {\it Bond number} which measures the relative strength of 
surface tension and the gravitational force.
%

Keeping only the first two terms in (\ref{kdvoriginal}), the speed of the 
associated linear (long) wave is $c = \sqrt{gh}.$
This is indeed the maximum attainable speed of propagation of gravity-induced 
water waves of infinitesimal amplitude.
The speed of propagation of the small-amplitude solitary waves described by
(\ref{kdvoriginal}) is slightly higher. 
According to Russell's empirical formula the speed equals 
$\sqrt{g (h + k)},$ where $k$ is the height of the peak of the solitary wave
above the surface of undisturbed water.
As Bullough (1988) has shown, Russell's approximate speed and the true speed 
of solitary waves only differ by a term of $O(k^2/h^2).$

The KdV equation can be recast in dimensionless variables as 
\begin{equation}
\label{kdvdimless}
u_t + \alpha u u_x + u_{xxx} = 0, 
\end{equation}
where subscripts denote partial derivatives.
The parameter $\alpha$ can be scaled to any real number. 
Commonly used values are $\alpha = \pm 1$ or $\alpha = \pm 6.$ 

The term $u_t$ describes the time evolution of the wave propagating in one 
direction.
Therefore, (\ref{kdvdimless}) is called an {\it evolution} equation. 
The nonlinear term $\alpha u u_x$ accounts for steepening of the wave, 
and the linear dispersive term $u_{xxx}$ 
describes spreading of the wave. 
The linear first-order term $\sqrt{gh} \frac{\partial \eta}{\partial X}$ 
in (\ref{kdvoriginal}) can be removed by an elementary transformation.
Conversely, a linear term in $u_x$ can be added to (\ref{kdvdimless}).

The nonlinear steepening of the water wave can be balanced by dispersion. 
If so, the result of these counteracting effects is a stable solitary wave 
with particle-like properties.
A solitary wave has a finite amplitude and propagates at constant speed and
without change in shape over a fairly long distance.
This is in contrast to the concentric group of small-amplitude capillary 
waves, shown in Figure~{\ref{sww-fig1-raindrops1}, which disperse as they 
propagate. 

The closed-form expression of a solitary wave solution is given by 
\begin{eqnarray}
\label{kdvsolitary1}
u(x,t) &=& \frac{\omega - 4 k^3}{\alpha k} 
+ \frac{12 k^2}{\alpha} {\sech}^2 (k x - \omega t + \delta) 
\\
\label{kdvsolitary2}
&=& \frac{\omega + 8 k^3}{\alpha k} 
- \frac{12 k^2}{\alpha} {\tanh}^2 (k x - \omega t + \delta), 
\end{eqnarray}
where the wave number $k$, the angular frequency $\omega,$ and $\delta$ are 
arbitrary constants. 

Requiring that $\lim_{x \to \pm \infty} u(x,t) = 0$ for all time leads to 
$\omega = 4 k^3.$ 
Then (\ref{kdvsolitary1}) and (\ref{kdvsolitary2}) reduce to 
\begin{equation}
\label{kdvsolitarysimpler}
u(x,t) = \frac{12 k^2}{\alpha} {\sech}^2 (k x - 4 k^3 t + \delta) 
= \frac{12 k^2}{\alpha} [1 - {\tanh}^2 (k x - 4 k^3 t + \delta) ]. 
\end{equation}
The position of the hump-type wave at $t = 0$ is depicted in 
Figure~{\ref{sww-fig5-sechcncolor}} for $\alpha = 6,$ $k = 2,$ and 
$\delta = 0.$ 
As time changes, the solitary wave with amplitude $2 k^2 = 8$ travels 
to the right at speed $v = \omega/k = 4 k^2 = 16.$ 
The speed is exactly twice the peak amplitude.
So, the taller the wave the faster it travels, but it does so without 
change in shape. 
The reciprocal of the wavenumber $k$ is a measure of the width of the 
sech-squared pulse.

\begin{figure}[htb]
\begin{center}
\includegraphics[width=4.7in, height=2.5in]{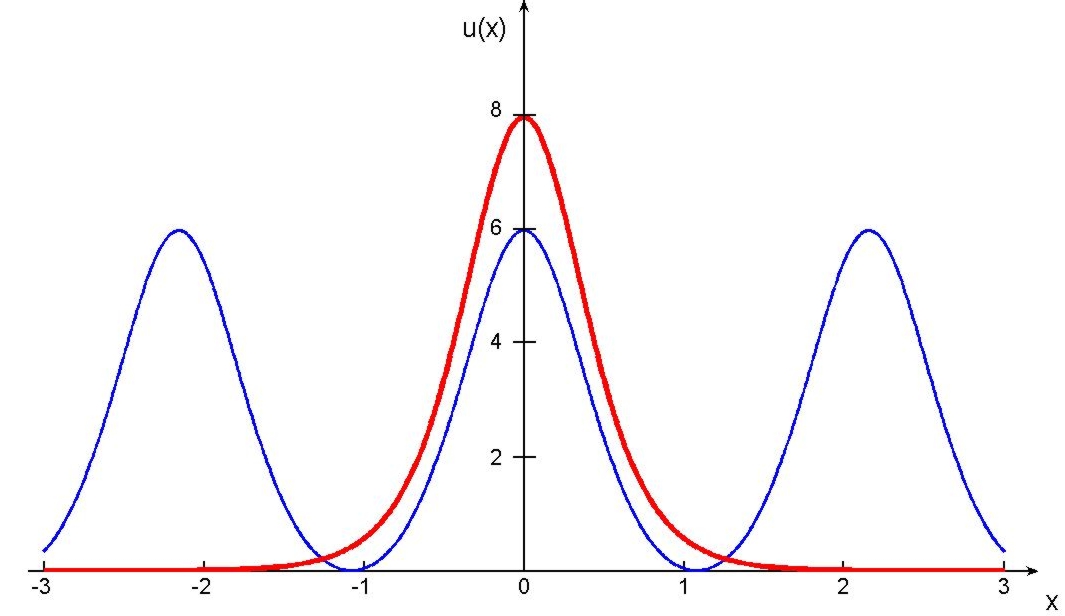}
\end{center}
\vspace{-6mm}
\caption{Solitary wave (red) and periodic cnoidal (blue) wave profiles.}
\label{sww-fig5-sechcncolor}
\end{figure}

As shown by Korteweg and de Vries (1895), equation (\ref{kdvdimless}) also 
has a simple periodic solution,
\begin{equation}
\label{kdvcnoidal}
u(x,t) = \frac{\omega - 4 k^3 (2 m - 1)}{\alpha k} 
+ \frac{12 k^2 m}{\alpha} {\cn}^2 (k x - \omega t + \delta ; m),
\end{equation}
which they called the {\it cnoidal wave} solution for it involves the 
Jacobi elliptic cosine function, $\cn,$ with modulus $m \, (0 < m < 1).$
%
The wavenumber $k$ gives the characteristic width of each oscillation in the 
``cnoid."

Three cycles of the cnoidal wave are depicted in 
Figure~{\ref{sww-fig5-sechcncolor}} at $t = 0.$  
The graph corresponds to $\alpha = 6, k = 2, m = 3/4, \omega = 16,$ and 
$\delta = 0.$
Using the property $\lim_{m \rightarrow 1} \cn(\xi; m) = \sech(\xi),$ 
one readily verifies that (\ref{kdvcnoidal}) reduces to 
(\ref{kdvsolitary1}) as $m$ tends to 1. 
Pictorially, the individual oscillations then stretch infinitely far apart 
leaving a single-pulse solitary wave.

The celebrated KdV equation appears in all books and reviews about 
soliton theory.
In addition, the equation has been featured in, e.g., Miles (1981) and 
Miura (1976).


\subsection*{Regularized Long-Wave Equations}

A couple of alternatives to the KdV equation have been proposed. 
A first alternative, 
\begin{equation}
\label{bbm}
u_t + u_x + \alpha u u_x - u_{xxt} = 0, 
\end{equation}
was proposed by Benjamin, Bona, and Mahony (1972).
Hence, (\ref{bbm}) is referred to as the BBM or regularized long-wave (RLW) 
equation. 

Equation (\ref{bbm}), which has a solitary wave solution,
\begin{equation}
\label{bbmsolitary}
u(x,t) = \frac{\omega - k - 4 k^2 \omega}{\alpha k} 
+ \frac{12 k \omega}{\alpha} {\sech}^2 (k x - \omega t + \delta),
\end{equation}
was also derived by Peregrine (1966) to describe the behavior of an undular 
bore (in water), which comprises a smooth wavefront followed by a train of 
solitary waves.
An undular bore can be interpreted as the dispersive analog of a shock wave 
in classical non-dispersive, dissipative hydrodynamics (El 2007).
%

The linear dispersion relation for the KdV equation, $\omega = k(1 - k^2),$
can be obtained by substituting $u(x,t) = \exp[i (k x - \omega t + \delta)]$ 
into $u_t + u_x + u_{xxx} = 0.$ 
The linear phase velocity, $ v_p = \omega/k = 1 - k^2,$ becomes negative for 
$|k|>1,$ thereby contradicting the assumption of uni-directional propagation.
Furthermore, the group velocity $v_g = d\omega/dk = 1 - 3 k^2$ has no 
lower bound which implies that there is no limit to the rate at which 
shorter ripples propagate in the negative $x-$direction.

The BBM equation, where
$\omega = k/(1 + k^2), v_p = 1/( 1 + k^2),$ and $v_g = (1-k^2)/(1+k^2)^2,$
was proposed to get around these shortcomings and to address issues related 
to proving the existence of solutions of the KdV equation.
The dispersion relation of (\ref{bbm}) has more desirable properties for 
high wave numbers, but the group velocity becomes negative for $|k|>1.$ 
In addition, the KdV and BBM equations are first order in time making it 
impossible to specify initial data for both $u$ and $u_t.$ 

To circumvent these limitations, a second alternative,
\begin{equation}
\label{josephegri}
u_t + u_x + \alpha u u_x + u_{xtt} = 0, 
\end{equation}
was proposed by Joseph and Egri (1977) and Jeffrey (1978).
It is called the time regularized long-wave (TRLW) equation and its 
solitary wave solution is given by 
\begin{equation}
\label{josehpsolitary}
u(x,t) = \frac{\omega - k - 4 k \omega^2}{\alpha k} 
+ \frac{12 \omega^2}{\alpha} {\sech}^2 (k x - \omega t + \delta).
\end{equation}
The TRLW equation shares many of the properties of both the KdV and 
BBM equations, at the cost of a more complicated dispersion relation, 
$\omega = (- 1 \pm \sqrt{1 + 4 k^2})/2 k$ with two branches.
Only one of these branches is valid because the derivation of the TRLW 
equation shows that (\ref{josephegri}) is uni-directional, despite the
presence of two time derivatives in $u_{xtt}.$

Bona and Chen (1999) have shown that the initial value problem for the TRLW 
equation is well-posed, and that for small-amplitude, long waves, 
solutions of (\ref{josephegri}) agree with solutions of either 
(\ref{kdvdimless}) or (\ref{bbm}). 
As a matter of fact, all three equations agree to the neglected order of 
approximation over a long time scale, provided the initial data is properly 
imposed (see also Bona {\it et al.} 1981).
%

Fine-tuning the dispersion relation of the KdV equation comes at a cost.
In contrast to (\ref{kdvdimless}), the RLW and TRLW equations are no 
longer completely integrable. 
Perhaps that is why these equations never became as popular as the KdV 
equation.
\subsection*{The Boussinesq Equation}
The classical Boussinesq equation,   
\begin{equation}
\label{boussinesqoriginal}
\eta_{TT} - c^2 \eta_{XX} - \frac{3 c^2}{h} (\eta_X^2 + \eta \eta_{XX}) 
- \frac{c^2 h^2}{3} \eta_{XXXX} = 0, 
\end{equation}
was derived by Boussinesq (1871) to describe gravity-induced surface waves 
as they propagate at constant (linear) speed $c = \sqrt{gh}$ in a canal of 
uniform depth $h.$

In contrast to the KdV equation, (\ref{boussinesqoriginal}) has a 
second-order time-derivative term. 
Ignoring all but the first two terms in (\ref{boussinesqoriginal}), 
one obtains the linear {\it wave} equation, $\eta_{TT} - c^2 \eta_{XX} = 0,$
which describes both left-running and right-running waves. 
However, (\ref{boussinesqoriginal}) is {\it not} bi-directional because in 
the derivation Boussinesq used the constraint $\eta_T = - c \eta_X,$ 
which limits (\ref{boussinesqoriginal}) to waves travelling to the right.
This crucial restriction if often overlooked in the literature.
%
%

In dimensionless form, the Boussinesq equation reads 
\begin{equation}
\label{boussinesqdimless}
u_{tt} - c^2 u_{xx} - \alpha u_x^2 - \alpha u u_{xx} - \beta u_{xxxx} = 0.
\end{equation}
The values of the parameters $c, \alpha > 0,$ and $\beta$ do not matter,
but the sign of $\beta$ matters. 
Typically, one sets $c = 1, \alpha = 3,$ and $\beta = \pm 1.$ 

A simple solitary wave solution of (\ref{boussinesqdimless}) is given by 
\begin{equation}
\label{boussinesqsolitary}
u(x,t) = \frac{\omega^2 - c^2 k^2 - 4 \beta k^4}{\alpha k^2} 
+ \frac{12 \beta k^2}{\alpha} {\sech}^2 (k x - \omega t + \delta). 
\end{equation}

The equation with $\beta = 1$ is a scaled version of 
(\ref{boussinesqoriginal}) and thus most relevant to shallow 
water wave theory. 
Mathematically, (\ref{boussinesqdimless}) with $\beta =1$ is ill-posed,
even without the nonlinear terms, which means that the initial value 
problem cannot be solved for arbitrary data.
This shortcoming does not happen for (\ref{boussinesqdimless}) with 
$\beta = -1,$ which is therefore nicknamed the ``good" Boussinesq 
equation (McKean 1981). 
Nonetheless, the classical and good Boussinesq equations are completely 
integrable.

The ``improved" or ``regularized" Boussinesq equation 
(see, e.g., Bona {\it et al} 2002)
has $\beta = 1$ but $u_{xxtt}$ instead of 
$u_{xxxx},$ which improves the properties of the dispersion relation.
Like (\ref{boussinesqdimless}), the regularized version describes 
uni-directional waves.
The regularized Boussinesq equation and other alternative equations 
listed in the literature (see, e.g., Madsen and Sch\"affer 1999) are not 
completely integrable.

%
Bona {\it et al} (2002, 2004) analyzed a family of Boussinesq systems 
of the form
\begin{eqnarray}
\label{boussinesqfamily}
w_t + u_x + (u w)_x + \alpha u_{xxx} - \beta w_{xxt} &=& 0, 
\nonumber \\
u_t + w_x + u u_x + \gamma w_{xxx} - \delta u_{xxt} &=& 0, 
\end{eqnarray}
which follow from the Euler equations as first-order approximations in 
the parameters $\epsilon_1 = a/h << 1, \epsilon_2 = h^2/\lambda^2 << 1,$ 
where the Stokes number, 
$S = \epsilon_1/\epsilon_2 = a \lambda^2/h^3 \approx 1.$

In (\ref{boussinesqfamily}) $w(x,t)$ is the nondimensional deviation 
of the water surface from its undisturbed position;
$u(x,t)$ is the nondimensional horizontal velocity field at a height 
$\theta h$ (with $0 \le \theta \le 1)$ above the flat bottom of the water.
The constant parameters $\alpha$ through $\delta$ in (\ref{boussinesqfamily})  
satisfy the following consistency conditions:
$\alpha + \beta = \frac{1}{2}(\theta^2 - \frac{1}{3})$ and 
$\gamma + \delta = \frac{1}{2}(1 - \theta^2) \ge 0.$ 
Solitary wave solutions of various special cases of 
(\ref{boussinesqfamily}) have been computed by Chen (1998). 
%
%

Boussinesq systems arise when modeling the propagation of long-crested 
waves on large bodies of water (such as large lakes or the ocean).
The Boussinesq family (\ref{boussinesqfamily}) encompasses many systems 
that appeared in the literature.
Special cases and properties of well-posedness of (\ref{boussinesqfamily}) 
are addressed by Bona {\it et al} (2002, 2004).

\subsection*{1D Shallow Water Wave Equation}
The so-called one-dimensional (1D) shallow water wave equation,
\begin{equation}
\label{1dsww}
v_{xxt} + \alpha v v_t - v_{t} - v_{x} 
+ \beta v_{x} \int_{\infty}^{x} v_t(y,t) dy = 0, 
\end{equation}
can be derived from classical shallow water wave theory (see Section IV) 
in the Boussinesq approximation.
In that approximation one assumes that vertical variations of the static 
density, $\rho_0,$ are neglected, except the buoyancy term proportional to 
$d\rho_0/dz,$ which is, in fact, responsible for the existence of solitary 
waves.
The integral term in (\ref{1dsww}) can be removed by introducing the
potential $u.$
Indeed, setting $v = u_x,$ equation (\ref{1dsww}) can be written as
\begin{equation}
\label{1dswwpot}
u_{xxxt} + \alpha u_{x} u_{xt} - u_{xt} - u_{xx} + \beta u_{xx} u_t  = 0.
\end{equation}
The equation is completely integrable and can be solved with the IST 
if and only if either $\alpha = \beta$ (Hirota and Satsuma 1976)
or $\alpha = 2 \beta$ (Ablowitz {\it et al} 1974).
When $\alpha = \beta,$ equation (\ref{1dswwpot}) can be integrated with
respect to $x$ and thus replaced by 
\begin{equation}
\label{1dswwhirota}
u_{xxt} + \alpha u_{x} u_{t} - u_{t} - u_{x} = 0.
\end{equation}
Closed-form solutions of (\ref{1dsww}), and in particular of 
(\ref{1dswwhirota}), have been computed by Clarkson and Mansfield (1994).
\subsection*{The Camassa-Holm Equation}
The CH equation, named after Camassa and Holm (1993, 1994), 
\begin{equation}
\label{camassaholm}
u_{t} + 2 \kappa u_{x} + 3 u u_x - \alpha^2 u_{xxt}  + \gamma u_{xxx}
- 2 \alpha^2 u_x u_{xx} - \alpha^2 u u_{xxx} = 0, 
\end{equation}
also models waves in shallow water.
In (\ref{camassaholm}), $u$ is the fluid velocity in the $x-$direction or, 
equivalently, the height of the water's free surface above a flat bottom, 
and $\kappa, \gamma$ and $\alpha$ are constants. 
Retaining only the first four terms in (\ref{camassaholm}) gives the 
BBM equation (\ref{bbm}).
Setting $\alpha = 0$ reduces (\ref{camassaholm}) to the KdV equation.

The CH equation admits solitary wave solutions, but in contrast to the 
hump-type solutions of the KdV and Boussinesq equations, they are 
implicit in nature (see, e.g., Johnson 2003).
In the limit $\kappa \rightarrow 0,$ equation (\ref{camassaholm}) with 
$\gamma = 0, \alpha = 1$ has a cusp-type solution of the form 
$u(x,t) = c \exp(-|x - c t - x_0|).$ 
The solution is called a peakon since it has a peak (or corner) where the
first derivatives are discontinuous.
The solution travels at speed $c > 0$ which equals the height of the peakon.

\subsection*{The Kadomtsev-Petviashvili Equation}
%
In their 1970 study of the stability of line solitons, Kadomtsev and 
Petviashvili (KP) derived a 2D-generalization of the KdV equation which 
now bears their names.
In dimensionless variables, the KP equation is 
\begin{equation}
\label{kpdimless}
(u_t + \alpha u u_x + u_{xxx})_x + \sigma^2 u_{yy} = 0, 
\end{equation}
where $y$ is the transverse direction.
In the derivation of the KP equation, one assumes that the scale of 
variation in the $y-$direction 
which is along the crest of the wave 
(as shown in Figure~{\ref{sww-fig4-testwave2}}) 
is much longer than the wavelength along the $x-$direction.

The solitary wave and periodic (cnoidal) solutions of (\ref{kpdimless}) 
are, respectively, given by 
\begin{equation}
\label{kpsolitary}
u(x,t) = \frac{k \omega - 4 k^4 + \sigma^2 l^2}{\alpha k^2} 
+ \frac{12 k^2}{\alpha} {\sech}^2 (k x + l y - \omega t + \delta), 
\end{equation}
and
\begin{equation}
\label{kpcnoidal}
u(x,t) = \frac{k \omega - 4 k^4 (2 m - 1) - \sigma^2 l^2}{\alpha k^2} 
+ \frac{12 k^2 m}{\alpha} {\cn}^2 (k x + l y - \omega t + \delta ; m).
\end{equation}

As shown in Figure~{\ref{sww-fig6-coastlima}}, near a flat beach the 
periodic waves appear as long, quasilinear stripes with a cn-squared cross 
section. 
Such waves are typically generated by wind and tides.
\vskip 0.60cm
\noindent
\begin{figure}[htb]
\begin{center}
\includegraphics[width=3.5in, height=3.0in]{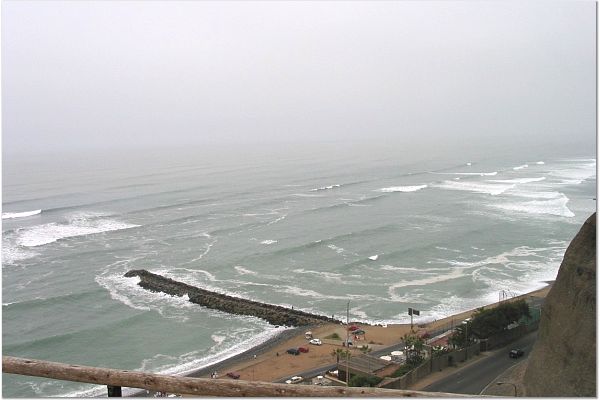}
\end{center}
\caption{Periodic plane waves in shallow water, off the coast of Lima, Peru. 
Photograph courtesy of A.\ Segur.}
\label{sww-fig6-coastlima}
\end{figure}

The equation with $\sigma^2 = -1$ is referred to as KP1, 
whereas (\ref{kpdimless}) with $\sigma^2 = 1$ is called KP2,
which describes shallow water waves (Segur 2007a).
Both KP1 and KP2 are completely integrable equations but their solution 
structures are fundamentally different (see, e.g., Scott 2005, pp.\ 489-490).
\section*{IV.$\;$Shallow Water Wave Equations of Geophysical Fluid Dynamics}
The shallow water equations used in geophysical fluid dynamics are 
based on the assumption 
$D/L << 1,$ 
where $D$ and $L$ are characteristic values for the vertical and horizontal 
length scales of motion.
For example, $D$ could be the average depth of a layer of fluid (or the 
entire fluid) and $L$ could be the wavelength of the wave. 

\begin{figure}[htb]
\begin{center}
%
%
\includegraphics[width=3.25in, height=2.75in]{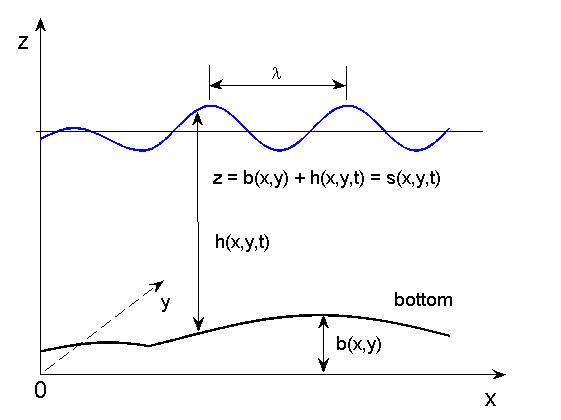}
\end{center}
\caption{Setup for the geophysical shallow water wave model.}
\label{sww-fig7-testwave1}
\end{figure}

The geophysical fluid dynamics community 
(see, e.g., Pedlosky 1987, Toro 2001, Vallis 2006) uses the following 
2D shallow water equations,  
\begin{eqnarray}
\label{swwmomentum1}
u_t + u u_x + v u_y  + g h_x - 2 \Omega v &=& - g b_x, 
\\
\label{swwmomentum2}
v_t + u v_x + v v_y + g h_y + 2 \Omega u &=& - g b_y,  
\\
\label{swwmass1}
h_t + (h u)_x + (h v)_y &=& 0,
\end{eqnarray}
to describe water flows with a free surface under the influence of gravity 
(with gravitational acceleration $g)$ and the Coriolis force due to the 
earth's rotation (with angular velocity $\Omega.)$
As usual, ${\bf u} = (u,v)$ denotes the horizontal velocity of the fluid 
and $h(x,y,t)$ is its depth. 
As shown in Figure~{\ref{sww-fig7-testwave1}}, $h(x,y,t)$ is the 
distance between the free surface $z = s(x,y,t)$ and the variable bottom 
$b(x,y).$  
Hence, $s(x,y,t) = b(x,y) + h(x,y,t).$
Equations (\ref{swwmomentum1}) and (\ref{swwmomentum2}) express the 
horizontal momentum-balance; (\ref{swwmass1}) expresses conservation of mass.
Note that the vertical component of the fluid velocity has been eliminated 
from the dynamics and that the number of independent variables has been 
reduced by one. 
Indeed, $z$ no longer explicitly appears in 
(\ref{swwmomentum1})-(\ref{swwmass1}), 
where $u,v,$ and $h$ only depend on $x,y,$ and $t.$ 

A shortcoming of the model is that it does not take into account the 
density stratification which is present in the atmosphere  
(as well as in the ocean).
Nonetheless, (\ref{swwmomentum1})-(\ref{swwmass1}) are commonly used by 
atmospheric scientists to model flow of air at low speed. 

More sophisticated models treat the ocean or atmosphere as a stack of 
layers with variable thickness. 
Within each layer, the density is either assumed to be uniform or may 
vary horizontally due to temperature gradients.
For example, Lavoie's rotating shallow water wave equations 
(see Dellar 2003), 
\begin{eqnarray}
\label{swwlavoievector}
{\bf u}_t + ({\bf u} {\bf \cdot} {\mbox {\boldmath $\nabla$}} ) {\bf u} 
   + 2 {\mbox{\boldmath $\Omega$}} \times {\bf u} 
   &=& - {\mbox {\boldmath $\nabla$}} (h \theta ) 
     + {\textstyle \frac{1}{2}} h {\mbox {\boldmath $\nabla$}} \theta,
\\
h_t + {\mbox {\boldmath $\nabla$}} {\bf \cdot} (h {\bf u}) &=& 0, 
\\
\theta_t + {\bf u} {\bf \cdot} ({\mbox {\boldmath $\nabla$}} \theta) &=& 0, 
\end{eqnarray}
consider only one active layer with layer depth $h(x,y,t),$ but take into 
account the forcing due to a horizontally varying potential temperature 
field $\theta(x,y,t).$
Vector ${\bf u} = u(x,y,t) {\bf i} + v(x,y,t) {\bf j}$ denotes the 
fluid velocity and ${\mbox {\boldmath $\Omega$}} = \Omega {\bf k}$ 
is the angular velocity vector of the Earth's rotation.
${\mbox {\boldmath $\nabla$}} = \frac{\partial}{\partial x} {\bf i} 
+ \frac{\partial}{\partial y} {\bf j}$ is the gradient operator, and 
${\bf i}$, ${\bf j},$ and ${\bf k}$ are unit vectors along the $x$, $y,$ 
and $z$-axes. 

Lavoie's equations are part of a family of multi-layer models proposed by 
Ripa (1993) to study, for example, the effects of solar heating, 
fresh water fluxes, and wind stresses on the upper equatorial ocean.
A study of the validity of various layered models has been presented 
by Ripa (1999).
Obviously, the more sophisticated the models become the harder they 
become to treat with analytic methods so one has to apply numerical methods.
Numerical aspects of various shallow water models in atmospheric research 
and beyond are discussed by, e.g., Weiyan (1992), Vreugdenhil (1994), and 
LeVeque (2002).

%

\section*{V.$\;$Computation of Solitary Wave Solutions}
\label{methods}
As shown in Section III,
solitary wave solutions of the KdV and Boussinesq equations 
(and like PDEs), can be expressed as polynomials of the hyperbolic 
secant (sech) or tangent (tanh) functions,
whereas their simplest period solutions involve the 
Jacobi elliptic cosine (cn) function.

There are several methods to compute exact, analytic expressions for 
solitary and periodic wave solutions of nonlinear PDEs. 
Two straightforward methods, namely the direct integration method and the 
tanh-method, will be discussed in this seciton.
Both methods seek travelling wave solutions. 
By working in a travelling frame of reference the PDE is replaced by an 
ordinary differential equation (ODE) for which one seeks closed-form 
solutions in terms of special functions.

In the terminology of dynamical systems, the solitary wave solutions
correspond to heteroclinic or homoclinic trajectories in the phase plane
of a first-order dynamical system corresponding to the 
underlying ODE (see, e.g., Balmforth 1995).
The periodic solutions (often expressible in terms of Jacobi elliptic 
functions) are bounded in the phase plane by these special trajectories, 
which correspond to the limit of infinite period and modulus one. 

Other more powerful methods, such as the forementioned IST and 
Hirota's method (see, e.g., Hirota 2004)
deal with the PDE directly.
These methods allow one to compute closed-form expressions of soliton  
solutions (in particular, solitary wave solutions) addressed elsewhere 
in the encyclopedia.
\vskip 5pt
\noindent
{\bf Direct Integration Method}
\vskip 4pt
\noindent
Exact expressions for solitary wave solutions can be obtained by direct 
integration. 
The steps are illustrated for the KdV equation given in (\ref{kdvdimless}).
Assuming that the wave travels to the right at speed $v = \omega/k,$ 
Equation (\ref{kdvdimless}) can be put into a travelling frame of 
reference with independent variable $\xi = k (x - v t - x_0).$ 
This reduces (\ref{kdvdimless}) to an ODE, 
$ - v \phi^{\prime} + \alpha \phi \phi^{\prime} 
+ k^2 \phi^{\prime\prime\prime} = 0, $
for $\phi(\xi) = u(x,t).$ 
A first integration with respect to $\xi$ yields 
\begin{equation}
\label{kdvodeint1}
- v \phi + \frac{\alpha}{2} \phi^2 + k^2 \phi^{\prime\prime} = A,
\end{equation}
where $A$ is a constant of integration. 
%
%
Multiplication of (\ref{kdvodeint1}) by $\phi^{\prime},$ followed by a 
second integration with respect to $\xi,$ yields 
\begin{equation}
\label{kdvodeint2}
- \frac{v}{2} \phi^2 + \frac{\alpha}{6} \phi^3 
+ \frac{k^2}{2} \phi^{\prime\, 2} = A \phi + B,
\end{equation}
where $B$ is an integration constant.
Separation of variables and integration then leads to 
\begin{equation}
\label{kdvodeint3}
\int_{\phi_0}^{\phi} 
\frac{d \phi}{\sqrt{a \phi^2 - b \phi^3 + \tilde{A} \phi + \tilde{B}}} 
= \pm \int_{\xi_0}^{\xi} d \xi, 
\end{equation}
where $a = v/k^2, b = \alpha/3 k^2, \tilde{A} = 2 A/k^2,$ and 
$ \tilde{B} = 2 B/k^2.$

The evaluation of the elliptic integral in (\ref{kdvodeint3}) depends 
on the relationship between the roots of the function 
$f(\phi) = a \phi^2 - b \phi^3 + \tilde{A} \phi + \tilde{B}.$
In turn, the nature of the roots depends on the choice of 
$\tilde{A}$ and $\tilde{B}.$ 
Two cases lead to physically relevant solutions. 
\vskip 2pt
\noindent
{\bf Case 1}: If the three roots are real and distinct, then the integral 
can be expressed in terms of the inverse of the $\cn$ function 
(see, e.g., Drazin and Johnson 1989 for details).
This leads to the cnoidal wave solution given in (\ref{kdvcnoidal}).
%
%
\vskip 2pt
\noindent
{\bf Case 2}: If the three roots are real and (only) two of them coincide, 
then the tanh-squared solution follows. 
This happens when $\tilde{A} = \tilde{B} = 0.$
Integrating both sides of (\ref{kdvodeint3}) then gives 
\begin{equation}
\label{kdvodeintegrated}
\int_{\phi_0}^{\phi} \frac{d \phi}{\phi \sqrt{a - b \phi}} 
= - \frac{2}{\sqrt{a}} \Arctanh[\frac{\sqrt{a - b \phi}}{\sqrt{a}}] + C
= \pm (\xi - \xi_0).
\end{equation}
where, without loss of generality, $C$ and $\xi_0$ can be set to zero.
Solving (\ref{kdvodeintegrated}) for $\phi$ yields
\begin{equation}
\label{phisol1}
\phi(\xi) = \frac{a}{b} (1 - {\tanh}^2 (\frac{\sqrt{a}}{2} \xi)) 
= \frac{a}{b} {\sech}^2(\frac{\sqrt{a}}{2} \xi).
\end{equation}
Returning to the original variables, one gets
\begin{equation}
\label{phisol2}
 \phi(\xi) 
= \frac{3 v}{\alpha} {\sech}^2 (\frac{\sqrt{v}}{2 k} \xi ), 
\end{equation}
or 
\begin{equation}
\label{kdvsolinverse}
u(x,t) = \frac{3 v}{\alpha} {\sech}^2 (\frac{\sqrt{v}}{2} (x - v t - x_0),
\end{equation} 
where $v$ is arbitrary.
Setting $v = \omega/k = 4 k^2,$
where $k$ is arbitrary, and 
$\delta = - k x_0,$ one can verify that (\ref{kdvsolinverse}) matches 
(\ref{kdvsolitarysimpler}).
\vskip 5pt
\noindent
{\bf The Tanh Method}
\vskip 4pt
\noindent
If one is only interested in tanh- or sech-type solutions, 
one can circumvent explicit integration 
(often involving elliptic integrals) 
and apply the so-called tanh-method.
A detailed description of the method has been given by 
Baldwin {\it et al} (2004). 
The method has been fully implemented in {\it Mathematica}, a popular 
symbolic manipulation program, and successfully applied to many
nonlinear differential equations from soliton theory and beyond. 

The tanh-method is based on the following observation: all derivatives of the 
$\tanh$ function can be expressed as polynomials in $\tanh.$
Indeed, using the identity ${\cosh}^2\xi - {\sinh}^2\xi = 1 $ one computes 
${\tanh}^{\prime}\xi = {\sech}^2\xi = 1 - {\tanh}^2\xi, \;$
${\tanh}^{\prime\prime}\xi = - 2 \, {\tanh}\,\xi + 2 \, {\tanh}^3\xi,$ etc. 
%
%
Therefore, all derivatives of $T(\xi) = {\tanh}\,\xi$ are polynomials in $T.$ 
For example, $T^{\prime} = 1 - T^2, T^{\prime\prime} = - 2 T + 2 T^3,$ and
$T^{\prime\prime\prime} = - 2 T + 8 T^2 - 6 T^4.$ 

By applying the chain rule, the PDE in $u(x,t)$ is then transformed into 
an ODE for $U(T)$ where 
$T = {\tanh}\,\xi = {\tanh}(k x - \omega t + \delta)$ is the new 
independent variable.
Since all derivatives of $T$ are polynomials of $T,$ the resulting ODE has 
polynomial coefficients in $T.$ 
It is therefore natural to seek a polynomial solution of the ODE.
The problem thus becomes algebraic.
Indeed, after computing the degree of the polynomial solution, one finds 
its unknown coefficients by solving a nonlinear algebraic system.

The method is illustrated using (\ref{kdvdimless}).
Applying the chain rule (repeatedly), the terms of (\ref{kdvdimless}) become 
$u_t = -\omega (1-T^2) U^{\prime},\; u_x = k (1-T^2) U^{\prime},$ and 
\begin{equation}
\label{kdvchainrule}
u_{xxx} =
k^3 (1 - T^2) \left[-2 (1 - 3 T^2) U^{\prime} - 6 T (1-T^2) U^{\prime\prime}
+ (1 - T^2)^2 U^{\prime\prime\prime} \right], 
\end{equation}
%
%
where $U(T) = U( \tanh(k x - \omega t + \delta) ) = u(x,t),$ 
$U^{\prime} = dU/dT,$ etc.

Substitution into (\ref{kdvdimless}) and cancellation of a common $1-T^2$ 
factor yields 
\begin{equation}
\label{kdvinU}
-\omega U^{\prime} 
+ \alpha k U U^{\prime} 
- 2 k^3 (1 - 3 T^2) U^{\prime}
- 6 k^3 T (1 - T^2) U^{\prime\prime}
+ k^3 (1 - T^2)^2 U^{\prime\prime\prime}  
= 0.
\end{equation}
This ODE for $U(T)$ has polynomial coefficients in $T.$
One therefore seeks a polynomial solution
\begin{equation}
\label{kdvpolsol}
U(T) = \sum_{n=0}^{N} a_n T^n,
\end{equation}
where the integer exponent $N$ and the coefficients $a_n$ must be computed.

Substituting $T^N$ into (\ref{kdvinU}) and balancing the highest powers 
in $T$ gives $N=2.$
Then, substituting 
\begin{equation}
\label{kdvsolU}
U(T) = a_0 + a_1 T + a_2 T^2
\end{equation} 
into (\ref{kdvinU}), and equating to zero the coefficients of the various 
power terms in $T,$ yields
\begin{eqnarray}
a_1 (\alpha a_2 + 2 k^2) &=& 0, 
\nonumber \\
\alpha a_2 + 12 k^2 &=& 0, 
\nonumber \\
a_1 (\alpha k a_0 - 2 k^3 - \omega) &=& 0, 
\\
\alpha k a_1^2  + 2 \alpha k a_0 a_2 - 16 k^3 a_2 - 2 \omega a_2 &=& 0. 
\nonumber
\end{eqnarray}
The unique solution of this nonlinear system for the unknowns $a_0, a_1,$ 
and $a_2$ is 
\begin{equation}
\label{kdvcoefs}
a_0 = \frac{8 k^3 + \omega}{\alpha k}, \;
a_1 = 0, \;
a_2 = -\frac{12 k^2}{\alpha}.
\end{equation}
Finally, substituting (\ref{kdvcoefs}) into (\ref{kdvsolU}) and 
using $T = \tanh(k x - \omega t + \delta )$ 
yields (\ref{kdvsolitary2}).

The solitary wave solutions and cnoidal wave solutions presented in 
Section III
have been automatically computed with a {\it Mathematica} package 
(Baldwin {\it et al} 2004) that implements the tanh-method and variants.

A review of numerical methods to compute solitary waves of arbitrary 
amplitude can be found in Vanden-Broeck (2007).
%
%
%
%
\section*{VI.$\;$Water Wave Experiments and Observations}
\label{experiments}
Through a series of experiments in a hydrodynamic tank, Hammack investigated 
the validity of the BBM equation (Hammack 1973) and KdV equation as models 
for long waves in shallow water (Hammack and Segur 1974, 1978a, 1978b) and 
long internal waves (Segur and Hammack 1982). 
Their research addressed the question: 
Would an initial displacement of water, as it propagates forward,
eventually evolve in a train of localized solitary waves (solitons) 
and an oscillatory tail as predicted by the KdV equation?
Based on the experimental data, they concluded that 
(i) the KdV dynamics only occurs if the waves travel over a long distance, 
(ii) a substantial amount of water must be initially displaced (by a piston)
to produce a soliton train, 
(iii) the water volume of the initial wave determines the shape of the 
leading wave in the wave train, and 
(iv) the initial direction of displacement (upward or downward piston motion) 
determines what happens later. 
Quickly raising the piston causes a train of solitons to emerge;
quickly lowering the piston causes all wave energy to distribute into the 
oscillatory tail, as predicted by the theory. 

Several other researchers have tested the validity of the KdV equation 
and variants in laboratory experiments 
(see, e.g., Remoissenet 1999, Helfrich and Melville 2006).
Bona {\it et al} (1981) give an in-depth evaluation of the BBM equation 
(\ref{bbm}) with and without dissipative term $u_{xx}.$
Their study includes 
(i) a numerical scheme with error estimates,
(ii) a convergence test of the computer code, 
(iii) a comparison between the predictions of the theoretical model and 
the results of laboratory experiments.
The authors note that it is important to include dissipative effects to 
obtain reasonable agreement between the forecast of the model and the 
empirical results.

Water tank experiments in conjunction with the analysis of actual data,
allows researchers to judge whether or not the KdV equation can be used 
to model the dynamics of tsunamis (see Segur 2007a). 
Tsunami research intensified after the December 2004 tsunami devastated 
large coastal regions of India, Indonesia, Sri Lanka, and Thailand, 
and killed nearly 300,000 people.

Apart from shallow water waves near beaches, the KdV equation and its 
solitary wave solution also apply to internal waves in the ocean.
Internal solitary waves in the open ocean are slow waves of large 
amplitude that travel at the interface of stratified layers of different 
density. 
Stratification based on density differences is primarily due to 
variations in temperature or concentration (e.g., due to salinity gradients). 
For example, absorption of solar radiation creates a near surface thin layer 
of warmer water (of lower density) above a thicker layer or colder, 
denser water. 
The smaller the density change, the lower the wave frequency, and the 
slower the propagation speed.
If the upper layer is thinner than the lower one, then the internal wave 
is a wave of depression causing a downward displacement of the fluid 
interface. 

Internal solitary waves are ubiquitous in stratified waters, in particular,
whenever strong tidal currents occur near irregular topography.
Such waves have been studied since the 1960s.
An early, well-documented case deals with internal waves in the Andaman Sea, 
where Perry and Schimke (1965) found groups of internal waves up to 80 m 
high and 2000 m long on the main thermocline at 500 m in 1500-m deep water.
%
%
Their measurements were confirmed by Osborne and Burch (1980) who showed 
that internal waves in the Andaman Sea are generated by tidal flows and can 
travel over hundreds of kilometers.

Strong internal waves can affect biological life and interfere with underwater
navigation.
Understanding the behavior of internal waves can aid in the design of 
offshore production facilities for oil and natural gas. 

The near-surface current associated with the internal wave locally modulates 
the height of the water surface.
Hence, the internal wave leaves a ``signature" or ``footprint" at the sea 
surface in the form of a packet of solitary waves (sometimes called current 
rips or tide rips).
These visual manifestations appear as long, quasilinear stripes in satellite 
imagery or photographs taken during space flights. 
Over 50 case studies and hundreds of images of oceanic internal waves can 
be found in ``An Atlas of Internal Solitary-like Waves and Their Properties" 
(Jackson 2004). 

Figure~{\ref{sww-fig8-gibraltar}} shows a photograph of three solitary waves 
packets which are the surface signature of internal waves in the 
Strait of Gibraltar. 
The photograph was taken from the Space Shuttle on October 11, 1984.
Spain is to the North, Morocco to the South.
Alternate solitary wave packets move toward the northeast or the southeast. 
The amplitude of these waves is of the order of 50 m; their wavelength
is in the range of 500-2000 m.
The separation between the packets is approximately 30 km. 
Waves of longer wavelengths and higher amplitudes have travelled the furthest.
The number of oscillations within each packet increases as time goes on.
Solitary wave packets can reach 200 km into the Western Mediterranean sea 
and travel for more than two days before dissipating. 
A in-depth study of solitary waves in the Strait of Gibraltar can be found 
in Farmer and Armi (1988).

\begin{figure}[htb]
\begin{center}
\includegraphics[width=4.3in, height=4.0in]{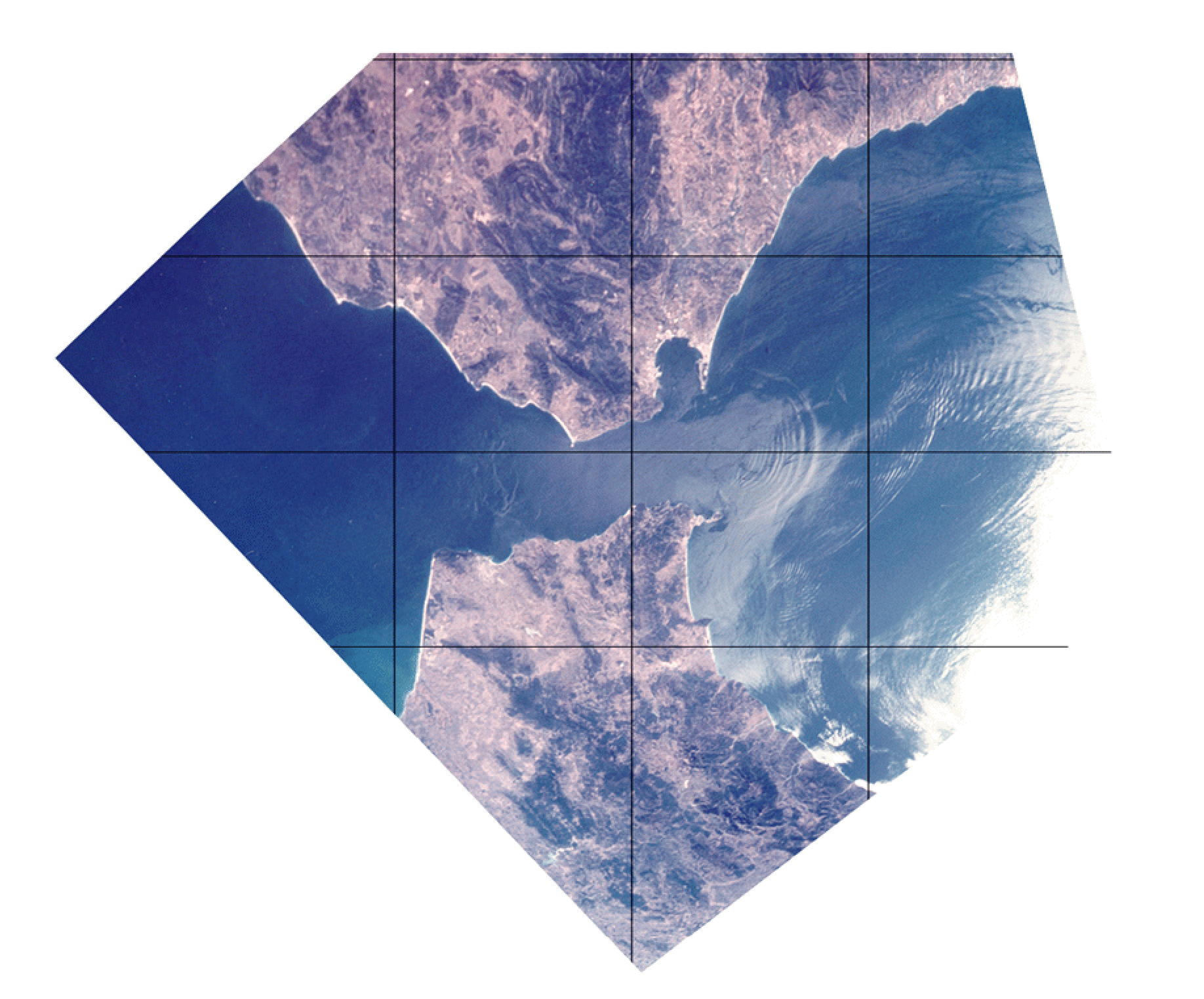}
\end{center}
\caption{Three solitary wave packets generated by internal waves from sills 
in the Strait of Gibraltar.
Original image STS41G-34-81 courtesy of the Earth Sciences and Image Analysis
Laboratory, NASA Johnson Space Center (http://eol.jsc.nasa.gov).
Ortho-rectified, color adjusted photograph courtesy of 
Global Ocean Associates.}
\label{sww-fig8-gibraltar}
\end{figure}
%
\vspace{-2mm}
\indent
The KdV model is applicable to stratified fluids with two layers and 
internal solitary waves if 
(i) the ratio of the amplitude $a$ to the upper layer depth $h$ is small, 
and 
(ii) the wavelength $\lambda$ is long compared with the upper layer depth.
More precisely, $a/h = O(h^2/\lambda^2) << 1.$ 
%
A detailed discussion of internal solitary waves and additional references 
can be found in Garrett and Munk (1979), Grimshaw (1997, 2001, 2005), 
Helfrich and Melville (2006), Apel {\it et al} (2007), and
Pelinovsky {\it et al} (2007).
The last three papers discuss a variety of other theoretical models 
including the extended KdV equation (also known as Gardner's equation or 
combined KdV-modified KdV equation) which contains both quadratic and 
cubic nonlinearities. 
Solitary wave solutions of the extended KdV equation can be found in 
Scott (2005, p.\ 856), Helfrich and Melville (2006), 
and Apel {\it et al} (2007).
A review of laboratory experiments with internal solitary waves was 
published by Ostrovsky and Stepanyants (2005).

As discussed in the review paper by Staquet and Sommeria (2002), 
internal gravity waves also occur in the atmosphere, where they are often 
caused by wind blowing over topography and cumulus convective clouds.
Internal gravity waves reveal themselves as unusual cloud patterns, 
which are the counterpart of the solitary wave packets on the ocean's surface.
%
%
%
%
\section*{VII.$\;$Future Directions}
\label{future}

For many shallow water wave applications, the full Euler equations are too 
complex to work with. 
Instead, various approximate models have been proposed.
Arguably, the most famous shallow water wave equations are the KdV and 
Boussinesq equations.

The KdV equation was originally derived to describe shallow water waves 
in a rectangular channel.
Surprisingly, the equation also models ion-acoustic waves and 
magneto-hydrodynamic waves in plasmas, waves in elastic rods,
mid-latitude and equatorial planetary waves, 
acoustic waves on a crystal lattice, 
and more (see, e.g., Scott {\it et al} 1973, and Scott 2003, 2005). 
The KdV equation has played a pivotal role in the development of the 
Inverse Scattering Transform and soliton theory, both of which had a 
lasting impact on twentieth-century mathematical physics.

Historically, the classical Boussinesq equation was derived to describe 
the propagation of shallow water waves in a canal. 
Boussinesq systems arise when modeling the propagation of long-crested 
waves on large bodies of water (e.g., large lakes or the ocean).
As Bona {\it et al} (2002) point out, a plethora of formally-equivalent 
Boussinesq systems can be derived. 
Yet, such systems may have vastly different mathematical properties.
The study of the well-posedness of the nonlinear models is of paramount 
importance and is the subject of ongoing research.

%
Shallow water wave theory allows one to adequately model waves in canals, 
surface waves near beaches, and internal waves in the ocean
(see Apel {\it et al} 2007).
Due to their widespread occurrence in the ocean (see Jackson 2004), 
solitary waves and ``solitary wave packets" (solitons) are of interest to 
oceanographers and geophysicists.
The (periodic) cnoidal wave solutions are used by coastal engineers in studies 
of sediment movement, erosion of sandy beaches, interaction of waves 
with piers and other coastal structures. 

Apart from their physical relevance, the knowledge of solitary and cnoidal
wave solutions of nonlinear PDEs facilitates the testing of numerical solvers
and also helps with stability analysis.

Shallow water wave models are widely used in atmospheric science as a 
paradigm for geophysical fluid motions. 
They model, for example, inertia-gravity waves with fast time scale dynamics, 
and wave-vortex interactions and Rossby waves associated with slow 
advective-timescale dynamics. 

This article has reviewed commonly used shallow water wave models, with the 
hope of bridging two research communities: one that focuses on 
nonlinear equations with dispersive effects; the other on nonlinear 
hyperbolic equations without dispersive terms. 
Of common concern are the testing of the theoretical models on measured data 
and further validation of the equations with numerical simulations and 
laboratory experiments.
A fusion of the expertise of both communities might advance research on water 
waves and help to answer open questions about wave breaking, instability, 
vorticity, and turbulence.
Of paramount importance is the prevention of natural disasters, ecological 
ravage, and damage to man-made structures due to a better understanding of 
the dynamics of tsunamis, steep waves, strong internal waves, rips, tidal 
currents, and storm surges. 
%
\vskip 15pt
\noindent
{\bf Acknowledgements}
\vskip 4pt
\noindent
%
This material is based upon work supported by the National Science Foundation
(NSF) of the USA under Award No.\ CCF-0830783. 
The author is grateful to Harvey Segur, Jerry Bona, Michel Grundland, 
and Douglas Poole for advise and comments.  
William Navidi, Benoit Huard, Anna Segur, Katherine Socha, 
Christopher Jackson, Michel Peyrard, and Christopher Eilbeck are 
thanked for help with figures and photographs.
%
%
%
%
\vskip 15pt
\noindent
{\sc Primary Literature}
\vskip 3pt
\noindent 
\begin{enumerate}
\item Ablowitz MJ, Clarkson PA (1991) 
Solitons, nonlinear evolution equations and inverse scattering.
Cambridge University Press, Cambridge, U.K.

\item Ablowitz MJ, Segur H (1981)
Solitons and the inverse scattering transform.
SIAM, Philadelphia, Pennsylvania

\item Ablowitz MJ et al (1974)
The inverse scattering transform: Fourier analysis for nonlinear problems.
Stud Appl Math 53:249-315

\item Apel JR et al (2007)
Internal solitons in the ocean and their effect on underwater sound.
J Acoust Soc Am 121:695-722


\item Baldwin D et al (2004)
Symbolic computation of exact solutions expressible in hyperbolic and 
elliptic functions for nonlinear PDEs.
J Symb Comp 37:669-705 

\item Balmforth NJ (1995)
Solitary waves and homoclinic orbits.
Annu Rev Fluid Mech 27:335-373

\item Benjamin TB et al (1972) 
Model equations for long waves in nonlinear dispersive systems.
Phil Trans Roy Soc London Ser A 272:47-78 

\item Bona JL, Chen H (1999) 
Comparison of model equations for small-amplitude long waves.
Nonl Anal 38:625-647

\item Bona JL et al (1981) 
An evaluation of a model equation for water waves.
Phil Trans Roy Soc London Ser A 302:457-510

\item Bona JL et al (2002) 
Boussinesq equations and other systems for small-amplitude long waves in 
nonlinear dispersive media. I: Derivation and linear theory.
J Nonl Sci 12:283-318 

\item Bona JL et al (2004) 
Boussinesq equations and other systems for small-amplitude long waves in 
nonlinear dispersive media. II: The nonlinear theory.
Nonlinearity 17:925-952.

\item Boussinesq J (1871)
Th\'eorie de l'intumescence liquide appel\'ee onde solitaire ou de 
translation, se propageant dans un canal rectangulaire.
C R Acad Sci Paris 72:755-759

\item Boussinesq J (1872) 
Th\'eorie des ondes et des remous qui se propagent le long d'un canal
rectangulaire horizontal, en communiquant au liquide contenu dans ce canal
des vitesses sensiblement pareilles de la surface au fond.
J Math Pures Appl 17:55-108

\item Boussinesq J (1877)
Essai sur la th\'eorie des eaux courantes.
Acad\'emie des Sciences de l'Institut de France, 
M\'emoires pr\'esent\'es par divers savants (ser 2) 23:1-680

\item Bullough RK (1988) 
``The wave" ``par excellence", the solitary, progressive great wave of 
equilibrium of the fluid--an early history of the solitary wave.
In: Lakshmanan M (ed) Solitons: introduction and applications.
Springer-Verlag, Berlin, pp 7-42

\item Camassa R, Holm D (1993) 
An integrable shallow water equation with peakon solitons. 
Phys Rev Lett 71:1661-1664 

\item Camassa R, Holm D (1994) 
An new integrable shallow water equation.
Adv Appl Mech 31:1-33 

\item Chen M (1998)
Exact solutions of various Boussinesq systems.
Appl Math Lett 11:45-49 

\item Clarkson PA, Mansfield EL (1994)
On a shallow water wave equation.
Nonlinearity 7:975-1000

\item Craik ADD (2004) 
The origins of water wave theory.
Annu Rev Fluid Mech 36:1-28 

\item Craik ADD (2005) 
George Gabriel Stokes and water wave theory.
Annu Rev Fluid Mech 37:23-42

\item Darrigol O (2003) 
The spirited horse, the engineer, and the mathematician: 
Water waves in nineteenth-century hydrodynamics.
Arch Hist Exact Sci 58:21-95

\item Dauxois T, Peyrard M (2004)
Physics of solitons.
Cambridge University Press, Cambridge, U.K.
Transl of Peyrard M, Dauxois T (2004)
Physique des solitons.
Savoirs Actuels, EDP Sciences, CNRS Editions


\item Dellar P (2003)
Common Hamiltonian structure of the shallow water equations
with horizontal temperature gradients and magnetic fields.
Phys Fluids 15:292-297

\item Drazin PG, Johnson RS (1989)
Solitons: an introduction.
Cambridge University Press, Cambridge, U.K.

\item El GA (2007)
Korteweg-de Vries equation: solitons and undular bores.
In: Grimshaw RHJ (ed) 
Solitary waves in fluids.
WIT Press, Boston, pp 19-53

\item Emmerson GS (1977)
John Scott Russell: A great Victorian engineer and naval architect.
John Murray Publishers, London


\item Farmer DM, Armi L (1988)
The flow of Mediterranean water through the Strait of Gibraltar.
Prog Oceanogr 21:1-105

\item Fermi F et al (1955)
Studies of nonlinear problems I.
Los Alamos Sci Lab Rep LA-1940
%
%
Reproduced in: AC Newell (ed) (1974) 
Nonlinear wave motion. 
AMS, Providence, Rhode Island

\item Filippov AT (2000) 
The versatile soliton.
Birkh\"auser-Verlag, Basel, Switzerland 

\item Garrett C, Munk W (1979)
Internal waves in the ocean.
Annu Rev Fluid Mech 11: 339-369

\item Grimshaw R (1997)
Internal solitary waves.
In: Liu PLF (ed) Advances in coastal and ocean engineering, vol III.
World Scientific, Singapore, pp 1-30

\item Grimshaw R (2001) 
Internal solitary waves.
In: Grimshaw R (ed) 
Environmental stratified flows.
Kluwer, Boston, pp 1-28

\item Grimshaw R (2005) 
Korteweg-de Vries equation.
In: Grimshaw R (ed) 
Nonlinear waves in fluids: Recent advances and modern applications.
Springer-Verlag, Berlin, pp 1-28

\item Hammack JL (1973) 
A note on tsunamis: their generation and propagation in an ocean of 
uniform depth. 
J Fluid Mech 60:769-800 

\item Hammack JL, Segur H (1974) 
The Korteweg-de Vries equation and water waves, part 2.
Comparison with experiments. 
J Fluid Mech 65:289-314 

\item Hammack JL, Segur H (1978a) 
The Korteweg-de Vries equation and water waves, part 3.
Oscillatory waves.
J Fluid Mech 84:337-358 

\item Hammack JL, Segur H (1978b) 
Modelling criteria for long water waves.
J Fluid Mech 84:359-373 

\item Helfrich KR, Melville WK (2006) 
Long nonlinear internal waves.
Annu Rev Fluid Mech 38:395-425

\item Hirota R (2004) 
The direct method in soliton theory.
Cambridge University Press, Cambridge, U.K.

\item Hirota R, Satsuma J (1976) 
N-soliton solutions of model equations for shallow water waves.
J Phys Soc Jpn 40:611-612 


\item Jackson CR (2004)
An atlas of internal solitary-like waves and their properties, 2nd ed. 
Global Ocean Associates, Alexandria, Virginia

\item Jeffrey A (1978) 
Nonlinear wave propagation.
Z Ang Math Mech (ZAMM) 58:T38-T56


\item Johnson RS (2003) 
On solutions of the Camassa-Holm equation.
Proc Roy Soc London Ser A 459:1687-1708 

\item Joseph RI, Egri R (1977) 
Another possible model equation for long waves in nonlinear dispersive systems.
Phys Lett A 61:429-432.

\item Kadomtsev BB, Petviashvili VI (1970)
On the stability of solitary waves in weakly dispersive media.
Sov Phys Doklady 15:539-541

\item Korteweg DJ, de Vries G (1895)
On the change of form of long waves advancing in a rectangular canal, and on 
a new type of long stationary waves.
Philos Mag (Ser 5) 39:422-443

\item LeVeque RJ (2002) 
Finite volume methods for hyperbolic problems.
Cambridge University Press, Cambridge, U.K.

\item Madsen PA, Sch\"affer HA (1999)
A review of Boussinesq-type equations for surface gravity waves.
In: Liu PLF (ed) Advances in coastal and ocean engineering, vol V.
World Scientific, Singapore, pp 1-95


\item McKean HP (1981) 
Boussinesq's equation on the circle.
Comm Pure Appl Math 34:599-691


\item Miles JW (1981)
The Korteweg-de Vries equation: a historical essay.
J Fluid Mech 106:131-147

\item Miura RM (1976)
The Korteweg-de Vries equation: a survey of results.
SIAM Rev 18:412-559

\item Osborne AR, Burch TL (1980) 
Internal solitons in the Andaman sea.
Science 208:451-460

\item Ostrovsky LA, Stepanyants YA (2005) 
Internal solitons in laboratory experiments: Comparison with theoretical 
models.
Chaos 15: 037111

\item Pedlosky J (1987)
Geophysical fluid dynamics, 2nd ed.
Springer-Verlag, Berlin


\item Pelinovsky E et al (2007)
In: Grimshaw RHJ (ed) 
Solitary waves in fluids.
WIT Press, Boston, pp 85-110

\item Peregrine DH (1966)
Calculations of the development of an undular bore.
J Fluid Mech 25:321-330 

\item Perry RB, Schimke GR (1965)
Large amplitude internal waves observed off the north-west coast of Sumatra.
J Geophys Res 70:2319-2324

\item Rayleigh L (1876)
On waves.
Philos Mag 1:257-279

\item Remoissenet M (1999)
Waves called solitons: Concepts and experiments, 3rd ed.
Springer-Verlag, Berlin

\item Ripa P (1993)
Conservation laws for primitive equations models with inhomogeneous layers.
Geophys Astrophys Fluid Dyn 70:85-111

\item Ripa P (1999)
On the validity of layered models of ocean dynamics and thermodynamics with 
reduced vertical resolution.
Dyn Atmos Oceans Fluid Dyn 29:1-40

\item Russell JS (1844)
Report on waves.
14th meeting of the British Association for the Advancement of Science, 
John Murray, London, pp 311-390

\item Scott AC (2003)
Nonlinear science: Emergence and dynamics of coherent structures.
Oxford University Press, Oxford, U.K.


\item Scott AC (ed) (2005) 
Encyclopedia of Nonlinear science.
Routledge, New York

\item Scott AC et al (1973)
The soliton: a new concept in applied science.
Proc IEEE 61:1443-1483


\item Segur H (2007a)
Waves in shallow water, with emphasis on the tsunami of 2004. 
In: Kundu A (ed) Tsunami and nonlinear waves,
Springer-Verlag, Berlin

\item Segur H (2007b)
Integrable models of waves in shallow water.
In: Pinski M and Birnir B (eds)
Probability, geometry and integrable systems.
Cambridge University Press, Cambridge, U.K.

\item Segur H, Hammack JL (1982)
Soliton models of long internal waves.
J Fluid Mech 118:285-304 

\item Staquet C, Sommeria J (2002) 
Internal gravity waves: From instabilities to turbulence.
Annu Rev Fluid Mech 34:559-593 

\item Stokes GG (1847)
On the theory of oscillatory waves.
Trans Camb Phil Soc 8:441-455

\item Toro EF (2001)
Shock-capturing methods for free-surface shallow flows.
Wiley, New York

\item Vallis GK (2006)
Atmospheric and oceanic fluid dynamics: fundamentals and large scale
circulation.
Cambridge University Press, Cambridge, U.K.

\item Vanden-Broeck J-M (2007)
Solitary waves in water: numerical methods and results.
In: Grimshaw RHJ (ed) 
Solitary waves in fluids.
WIT Press, Boston, pp 55-84

\item Vreugdenhil CB (1994)
Numerical methods for shallow-Water flow.
Springer-Verlag, Berlin

\item Weiyan T (1992)
Shallow water hydrodynamics: mathematical theory and numerical solution 
for two-dimensional systems of shallow water equations.
Elsevier, Amsterdam

\item Zabusky NJ (2005)
Fermi-Pasta-Ulam, solitons and the fabric of nonlinear and computational
science: History, synergetics, and visiometrics.
Chaos 15: 015102

\item Zabusky NJ, Kruskal MD (1965)
Interaction of `solitons' in a collisionless plasma and the recurrence
of initial states.
Phys Rev Lett 15:240-243

\end{enumerate}
%
\vskip 3pt
\noindent
{\sc Books and Reviews}
\vskip 3pt
\noindent
%
%
%
Boyd JP (1998)
Weakly nonlinear solitary waves and beyond-all-order asymtotics.
Kluwer, Dortrecht, The Netherlands
\vskip 3pt
\noindent
Calogero F, Degasperis A (1982)
Spectral transform and solitons I.
North Holland, Amsterdam
\vskip 3pt
\noindent
Dickey LA (2003)
Soliton equations and Hamiltonian systems, 2nd ed.
World Scientific, Singapore
\vskip 3pt
\noindent
Dodd RK et al (1982)
Solitons and nonlinear wave equations.
Academic Press, London
\vskip 3pt
\noindent
%
Eilenberger G (1981) 
Solitons: Mathematical methods for physicists.
Springer-Verlag, Berlin
\vskip 3pt
\noindent
Faddeev LD, Takhtajan LA (1987) 
Hamiltonian methods in the theory of solitons.
Springer-Verlag, Berlin
\vskip 3pt
\noindent
%
Fordy A (ed) (1990) 
Soliton theory: A survey of results.
Manchester University Press, Manchester
\vskip 3pt
\noindent
%
%
Grimshaw RHJ (ed) (2007)
Solitary waves in fluids.
WIT Press, Boston
\vskip 3pt
\noindent
%
Infeld E, Rowlands G (2000)
Nonlinear waves, solitons, and chaos
Cambridge University Press, New York 
\vskip 3pt
\noindent
Johnson RS (1977)
A modern introduction to the mathematical theory of water waves.
Cambridge University Press, Cambridge, U.K.
\vskip 3pt
\noindent
Kasman A (2010) Glimpses of soliton theory, AMS, Providence, Rhode Island
\vskip 3pt
\noindent
%
%
Lamb GL (1980)
Elements of soliton theory.
Wiley Interscience, New York
\vskip 3pt
\noindent
Miles JW (1980)
Solitary waves.
Annu Rev Fluid Mech 12:11-43
\vskip 3pt
\noindent
%
%
Mei CC et al (2005)
Theory and applications of ocean surface waves.
World Scientific, Singapore
\vskip 3pt
\noindent
Mitropol'sky YZ (2001)
Dynamics of internal gravity waves in the ocean.
Kluwer Academic, Dortrecht, The Netherlands
\vskip 3pt
\noindent
Newell AC (1983) 
The history of the soliton.
J Appl Mech 50:1127-1137 
\vskip 3pt
\noindent
Newell AC (1985) 
Solitons in mathematics and physics.
SIAM, Philadelphia, Pennsylvania
\vskip 3pt
\noindent
Makhankov VG (1990)
Soliton phenomenology.
Kluwer Academic, Dortrecht, The Netherlands
\vskip 3pt
\noindent
Novikov SP et al (1984)
Theory of solitons. The inverse scattering method.
Consultants Bureau, Plenum Press, New York
\vskip 3pt
\noindent
Osborne AR (2010) Nonlinear ocean waves \& the inverse scattering transform,
Academic Press, New York
\vskip 3pt
\noindent
Pedlosky J (2003)
Waves in the ocean and atmosphere: introduction to wave dynamics.
Springer-Verlag, Berlin
\vskip 3pt
\noindent
%
%
Russell JS (1885)
The wave of translation in the oceans of water, air and ether.
Tr\"ubner, London
\vskip 3pt
\noindent
Socha K (2007) Circles in circles. Creating a mathematical model of surface
water waves. MAA Monthly 114:202-216
\vskip 3pt
\noindent
Stoker JJ (1957)
Water waves.
Wiley Interscience, New York
\vskip 3pt
\noindent
Whitham GB (1974)
Linear and nonlinear waves.
Wiley Interscience, New York
\vskip 3pt
\noindent
Yang J (2010) Nonlinear waves in integrable and nonintegrable systems, SIAM,
Philadelphia, Pennsylvania
%
%
\end{document}